%% file: wAP.tex
\documentclass[journal]{IEEEtran}
\usepackage[ansinew]{inputenc}
\usepackage{graphicx}
\usepackage{epsfig}
\usepackage{amsmath}    % From the American Mathematical Society
\usepackage{amssymb}
\usepackage{subcaption}
\usepackage[all]{xy}
\usepackage{multirow}
\usepackage{tikz}
\usetikzlibrary{snakes,matrix,trees,arrows}

\usepackage[a4paper,left=1.8cm,right=1.8cm,top=2.4cm,bottom=2.4cm]{geometry}

\usepackage{float} %Para poder situar las figuras con H
\usepackage{color}
\usepackage{xcolor}
\usepackage{listings}

\makeatletter

\def\zapcolorreset{\let\reset@color\relax\ignorespaces}
\def\colorrows#1{\noalign{\aftergroup\zapcolorreset#1}\ignorespaces}

\makeatother

\begin{document}

\title{wARP-Path: Implications of adapting \\the Ethernet-based ARP-Path bridging protocol \\to a wireless environment}

\author{Elisa~Rojas,
		Hedayat~Hosseini,
		Andres~Beato,
		Jose~Manuel~Gimenez-Guzman
        and Guillermo~Ibanez% <-this % stops a space
\thanks{Elisa~Rojas, Guillermo~Ibanez and Jose~Manuel~Gimenez-Guzman are with Departamento de Automatica, University of Alcala, 28805, Alcala de Henares (Madrid), Spain. e-mails: elisa.rojas@uah.es, guillermo.ibanez@uah.es, josem.gimenez@uah.es}% <-this % stops a space
\thanks{Andres~Beato is with World Class Center of Advanced Networks, Altran Innovacion, Madrid, Spain. e-mail: andres.beatoollero@altran.com}% <-this % stops a space %TODO
\thanks{Hedayat~Hosseini is with Computer Engineering and Information Technology Department, Amirkabir University of Technology (Tehran Polytechnic), Tehran, Iran. e-mail: h.hosseini@aut.ac.ir}% <-this % stops a space %TODO
%\thanks{Manuscript received May 31, 2014.}
%\thanks{Revision submitted on October 6, 2014.}
}

\markboth{Version 1.00}%
{Rojas \MakeLowercase{\textit{et al.}}: wARP-Path: Implications of adapting the Ethernet-based ARP-Path bridging protocol to a wireless environment}

\maketitle

%\newpage %###

\begin{abstract}
The ARP-Path protocol has flourished as a promise for wired networks, creating shortest paths with the simplicity of pure bridging and competing directly with TRILL and SPB. After analyzing different alternatives of ARP-Path and creating the All-Path family, the idea of migrating the protocol to wireless networks appeared to be a good alternative to protocols such as a AODV. \\
In this article, we check the implications of adapting ARP-Path to a wireless environment, and we prove that good ideas for wired networks might not be directly applicable to wireless networks, as not only the media differs, but also the characterization of these networks varies. 
\end{abstract}


\IEEEpeerreviewmaketitle

\input{text/intro}
\input{text/related}

\input{text/fromto}

\input{text/implementation}

\input{text/evaluation}

\input{text/discussion}

\input{text/conclusion}

\section*{Acknowledgment}
This work has been supported by Comunidad de Madrid through project TIGRE5-CM (S2013/ICE-2919).

\ifCLASSOPTIONcaptionsoff
  \newpage
\fi

\bibliographystyle{IEEEtran}
\bibliography{wAP}   % name your BibTeX data base

\end{document}

%% file: text/intro.tex
\section{Introduction}
\label{introduction}
%What is the problem?\\
%Why is it interesting and important?\\
%Why is it hard? (E.g., why do naive approaches fail?)\\
%Why hasn't it been solved before? (Or, what's wrong with previous proposed solutions? How does mine differ?)\\
%What are the key components of my approach and results? Also include any specific limitations.\\} 
%From http://cs.stanford.edu/people/widom/paper-writing.html

%All-Path
The All-Path family~\cite{Rojas15} comprises diverse routing protocols running on layer 2, leveraging the well-known advantages of Ethernet in wired networks. All-Path protocols are based on the simple and basic mechanism of backward address learning used by bridges/switches, extended with a \textit{lock} mechanism to prevent loops. Paths are discovered and built based on minimum latency and may be created per destination host (ARP-Path), per communication flow or host pair (Flow-Path), per destination bridge (Bridge-Path) or even combining parameters from different layers, such as TCP (TCP-Path~\cite{tcppath}). 

%Ethernet competitors
The first protocol, and origin of the whole All-Path family, was ARP-Path~\cite{IbanezCL}, already implemented in multiple and diverse  platforms like Linux, NetFPGA, OpenFlow, or OMNeT++~\cite{Ibanez10}. Its main competitors are protocols using layer 2 variants of the IS-IS protocol such as SPB~\cite{Allan12} or TRILL RBridges~\cite{RBridges,Perlman11}. The approach for ARP-Path is to discover low latency paths via network exploration with broadcast frames, while in SPB and TRILL paths are computed based on the network topology obtained after an initial exchange of data at link level.

%Wireless networks
In its origins, the All-Path family was inspired by the Ad hoc On-Demand Distance Vector (AODV) routing~\cite{aodv,aodv-paper}, designed for wireless networks. Both create paths following a reactive routing approach. Therefore, one of the challenges was to adapt this family of bridging protocols from a wired environment to a wireless one, in order to analyze and compare them to AODV.

This paper is organized as follows. Chapter~\ref{introduction} introduces the topic, followed by the related work in Chapter~\ref{related}. Afterwards, Chapter~\ref{fromto} explains the implication of moving ARP-Path to a wireless environment and defines wARP-Path. Chapter~\ref{implementation} describes the implementation in the OMNeT++ simulator, later on evaluated in Chapter~\ref{evaluation}. Finally, Chapter~\ref{discussion} discusses the topic and Chapter~\ref{conclusion} concludes the article.

%% file: text/related.tex
\section{Related Work}
\label{related}

There is a vast literature related to routing protocols applied to wireless networks. More specifically, one of the most prominent research fields that has thoroughly studied this problem is the one linked to \textbf{ad hoc networks}, and more specially \textbf{wireless mesh networks (WMNs)}. A seminal paper related to routing strategies in wireless networks is~\cite{Iwata99}, where authors propose and evaluate two competitive proposals for routing, called fisheye state routing and hierarchical
state routing. However, probably the most well-known routing protocols for routing in wireless networks are Ad hoc On-Demand Distance Vector (AODV) and Dynamic Source routing (DSR), as they are both standardized by IETF. \\
\textbf{AODV} is described in RFC 3561~\cite{aodv} and is based in using destination sequence numbers to avoid loops even in such situations where there are anomalous delivery of routing control messages. To our purpose, it is also very interesting the work in~\cite{aodv-paper}, as it describes some of the most interesting evolutions of AODV that have improved issues like performance, robustness or scalability and sheds light on future evolutions for the protocols. In fact, AODV features are the grounds of the All-Path family protocols. \\ 
The second standardized protocol is \textbf{DSR} protocol and is defined in RFC 4728~\cite{dsr}. DSR is a distributed protocol able to work in multi-hop wireless networks and it discovers and maintains routes by means of two mechanisms: route discovery and route maintenance. More specifically, route discovery in DSR is based on source routing and route caches, maintaining multiple routes per destination. A performance comparison between AODV and DSR can be found in~\cite{Das00}.

Other routing protocols that focus on improving different issues have been proposed in the literature --since the proposal of AODV and DSR-- are Source Node Compute Routing (SNCR)~\cite{He10} or Protection AODV (P-AODV)~\cite{Zhu13}. In the first case, \textbf{SNCR} aims to improve overhead and efficiency proposing a quick computation of the best metric for arriving from the source node to any destination, combining proactive and on-demand modes to adapt to different traffic settings. On the other hand, \textbf{P-AODV} focuses on improving reliability in wireless networks by means of building a protection path. Another important issue in routing in wireless routing is related to the link quality evaluation. In this sense, it must be highlighted the work in~\cite{Hong15}, where authors propose a link quality prediction (LQP) model to sense the link state.

Finally, the comparison of \textbf{bridging and routing techniques in wireless networks} is also a topic aligned with our proposal.
In~\cite{Suliman04}, authors compare wireless bridging and routing, concluding that bridging performs better than routing in terms of throughput both in TCP and UDP. Moreover, authors in~\cite{Maurina10} study the consequences of using pure bridging based solutions in wireless networks and present an enhanced bridged-based implementation for providing dynamic, self-configuration and self-healing features avoiding a routing protocol.

%% file: text/fromto.tex
\section{From ARP-Path to wARP-Path}
\label{fromto}

In this section, we evaluate the key aspects for the transition from ARP-Path (wired and strictly Ethernet-based) to wARP-Path (wireless and potentially implemented using different layer-two protocols). For this purpose, we start by summarizing ARP-Path, then explaining the basics of wARP-Path (with emphasis on the differences), and finally analyze the implications of the frame format.

\subsection{ARP-Path}
\label{sec:ap}

The ARP-Path protocol creates minimum latency paths at request, based on path exploration instead of computation~\cite{IbanezCL,Rojas15}. ARP-Path does not require any modification of the Ethernet frame, and it only needs a small new feature in standard switches: a \textit{lock}. The \textit{lock} is a mechanism that prevents the switch from learning more than once (in a certain period of time) the same MAC address, which in the end prevents network routing loops~\cite{Rojas15} and allows any broadcast frame to explore the whole network as a probe, creating a source-based routing tree in its way.

\begin{figure}
        \centering
        \begin{subfigure}[htb]{0.48\textwidth}
                \includegraphics[width=\textwidth]{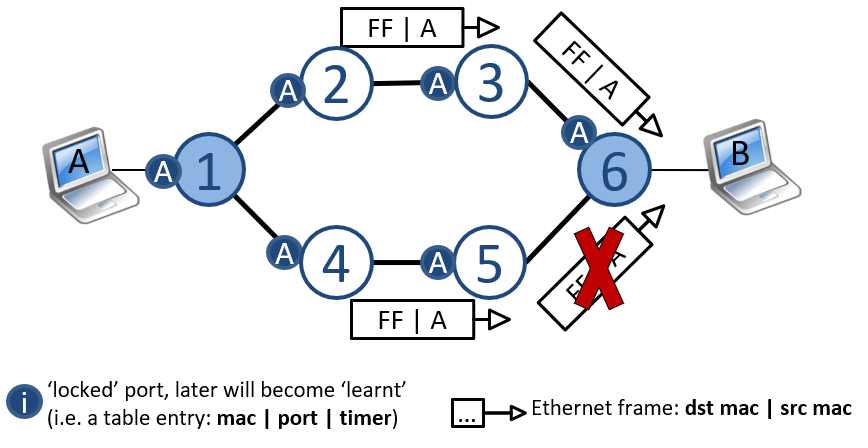}
                \caption{Learning process for path to $A$ when the ARP Request is broadcast}
                \label{ap1}
        \end{subfigure}%
        \quad
        \begin{subfigure}[htb]{0.48\textwidth}
                \includegraphics[width=\textwidth]{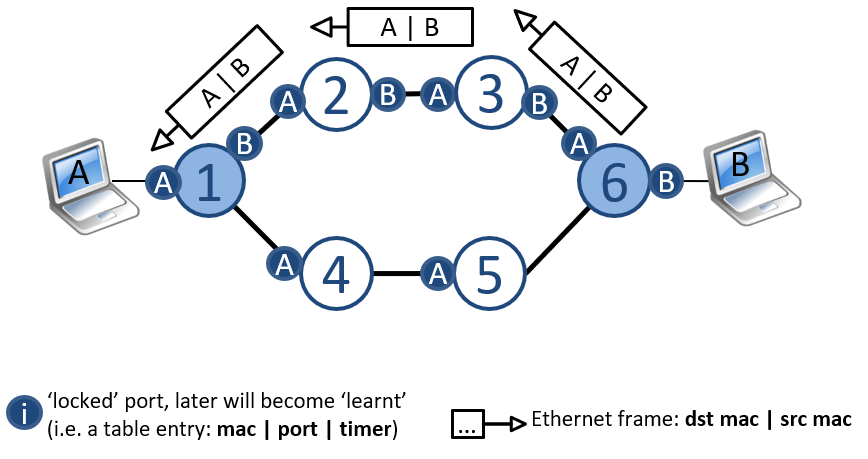}
                \caption{Learning process for path to $B$ when the ARP Reply is sent from $B$ to $A$}
                \label{ap2}
        \end{subfigure}
        \caption{ARP-Path operation in an Ethernet-based network}\label{fig:ap}
\end{figure}

Fig.~\ref{fig:ap} summarizes the operation of ARP-Path in an example Ethernet network that connects two hosts $A$ and $B$. Before any communication in IPv4, $A$ sends and ARP Request, which is leveraged to explore the topology and create the shortest paths. The difference with a standard switch is shown in switch 6, which only saves the first MAC address arriving --the one from switch 3-- and \textit{locks} the association of this MAC to the input port. Therefore, when the ARP Request arrives at a different port --from switch 5--, the frame is discarded, thus avoiding the potential loop, which would not happen in a standard Ethernet switch. Accordingly, only the fastest copy of the ARP Request arrives at $B$.

The learning process continues from $B$ to $A$ with the ARP Reply. This time, the unicast ARP Reply is forwarded towards $A$, already known by the different hops --switches-- in the network. Hence it crosses switches 6, 3, 2 and 1, respectively, until arriving at $A$. In the meantime, these switches save the input port for the ARP Reply, eventually generating the path towards $B$.

Although ARP-Path takes it name from the ARP protocol and bases its exploration in the ARP frames, it could be applied to non-ARP based networks, such as IPv6 networks. The only difference is that ARP-Path should use a specific frame for the exploration in that case.

\subsection{wARP-Path}
\label{sec:wap}

The wARP-Path protocol follows the same basics for creating paths to reach final hosts than ARP-Path. However, there are three main differences: 
\begin{enumerate}

\item \textbf{Locking mechanism:} The ARP-Path protocol locks the input port with the source MAC address in the arriving ARP message. In wireless networks there are no links and therefore no ports, so the locking mechanism saves the \texttt{\{MAC address, next hop's MAC address\}} tuple instead of the \texttt{\{MAC address, port\}} one. Basically, wARP-Path \textit{locks} nodes instead of input ports.
\item \textbf{Flooding:} To explore the network and reach the destination, the ARP-Path protocol broadcast frames through all ports but the input one. However, this concept is not directly applicable to wireless forwarding devices, which always broadcast frames in the whole radio coverage area.
\item \textbf{Forwarding nodes:} The ARP-Path protocol is used in bridge-based networks. However, wARP-Path can be applied to any ad hoc wireless network. For this reason, intermediate nodes can be final hosts at the same time, and therefore they might implement more communication layers than bridges (up to layer two only). This causes that ARP messages already have a default processing by intermediate nodes that should be slightly modified, i.e. intermediate nodes should discard only sARP messages not directed to them and not all of them (which is done in ARP-Path by default).
\end{enumerate}

\begin{figure}
        \centering
        \begin{subfigure}[htb]{0.48\textwidth}
                \includegraphics[width=\textwidth]{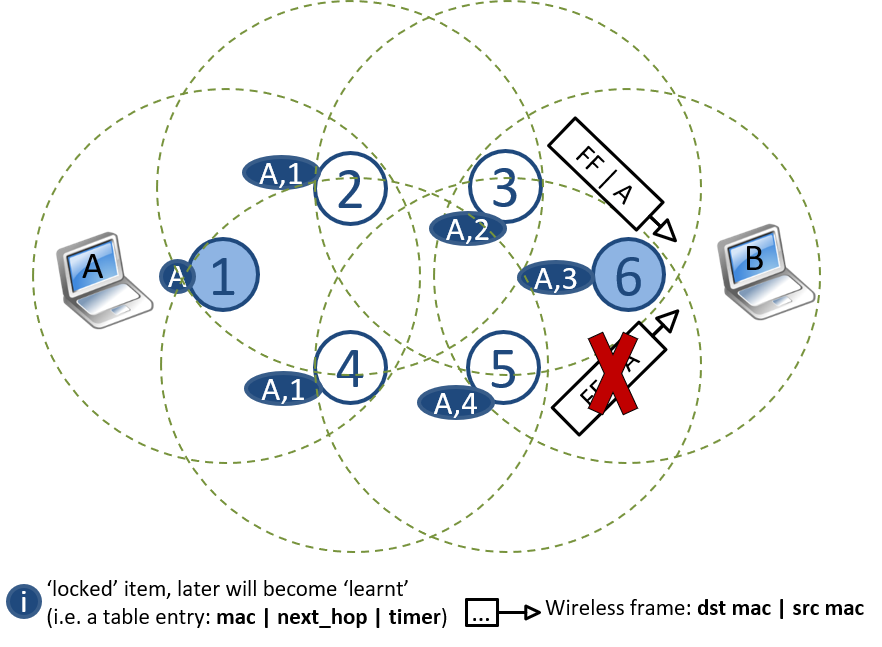}
                \caption{Learning process for path to $A$ when the ARP Request is broadcast}
                \label{wap1}
        \end{subfigure}%
        \quad
        \begin{subfigure}[htb]{0.48\textwidth}
                \includegraphics[width=\textwidth]{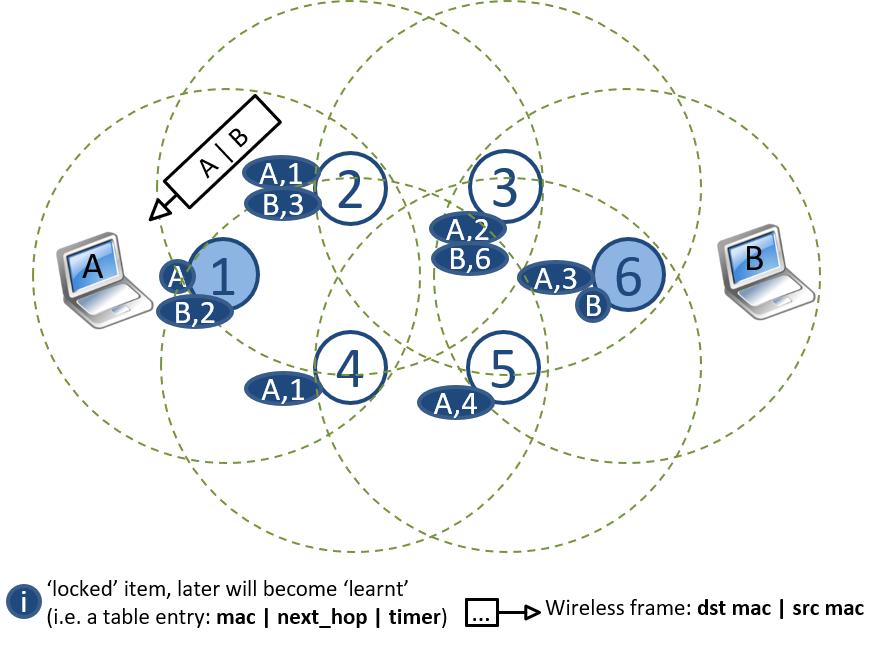}
                \caption{Learning process for path to $B$ when the ARP Reply is sent from $B$ to $A$}
                \label{wap2}
        \end{subfigure}
        \caption{wARP-Path operation in a wireless network}\label{fig:wap}
\end{figure}

Fig.~\ref{fig:wap} shows and example of a network, consisting of six intermediate nodes and two final hosts, in which wARP-Path can be applied. A circular dotted line represents the range or coverage area of each forwarding node. The range of final hosts is not represented, but we considered they reach only the closest node, for the sake of simplicity.

In this example, there are two possible paths between host $A$ and host $B$, same ones than in the example in Fig.~\ref{fig:ap}. 
When host $A$ emits the ARP Request message to start the communication, it first reaches node 1 which saves the tuple \texttt{A's MAC address, A's MAC address} since transmitter address of the frame received from host $A$ is regarded as the next hop address in host 1(i.e. in this case, the next hop is directly A). Later on, other nodes receive the frame, such as node 3, which saves the tuple \texttt{A's MAC address, 2's MAC address}, indicating that node 2 is locked as the next hop for the path to reach $A$. Eventually, several frames arrive to node 6, which only locks the MAC address of the first node from which it received the frame and discards the rest. 

Differently to ARP-Path, in wARP-Path many nodes receive back multiple copies of the frame and need to discard them. This is because wireless nodes emit in their range and cannot avoid \textit{emitting back} to the previous sender as in wired networks, where it is possible to flood through all ports but the incoming one. 

Finally, host $B$ receives the ARP Request message and emits the ARP Reply message with destination $A$. This ARP Reply message can follow the path just created to $A$ and, at the same time, it creates the path to $B$. Therefore the communication between $A$ and $B$ can start now, which will use the path involving nodes 1-2-3-6.

The pseudocode of the wARP-Path protocol is summarized in Listing~\ref{pseudocode}.

\begin{lstlisting}[basicstyle=\scriptsize, frame=single, caption={Pseudocode of the wARP-Path protocol}, label=pseudocode]
When a wARP-Path node receives a frame from another node:
01:if (src_mac ==  node's MAC address) then
02:    discard frame
03:else
04:    eth_frame = convertToEthernetFrame(frame)
05:    next_hop = wARP-Path table(eth_frame, input_hop)
06:    If (dst_mac ==  node's MAC address) then
07:        send frame to upper layers of the node
08:    else if (dst_mac ==  BCAST && next_hop) then
09:        send frame to upper layers of the node
10:    If (next_hop) then
11:        send frame to the next hop
12:    else
13:        discard frame
\end{lstlisting}

\subsection{From a wired to a wireless frame format }
\label{sec:frame}

Apart from the implications previously mentioned, the implementation of wARP-Path requires another one related to the frame format. \\
ARP-Path leverages the fact that Ethernet is the most commonly used layer 2 protocol, and reuses the Ethernet frame for its purpose. However, wARP-Path depends on the type of network that relies beneath and their corresponding frame format. 

Therefore, for the sake of simplicity, we considered wireless local area networks, following the standards defined by IEEE802.11~\cite{ieee80211}.

%% file: text/implementation.tex
\section{Implementation}
\label{implementation}

\begin{figure*}[hbt]
\centering
\includegraphics[width=0.9\textwidth]{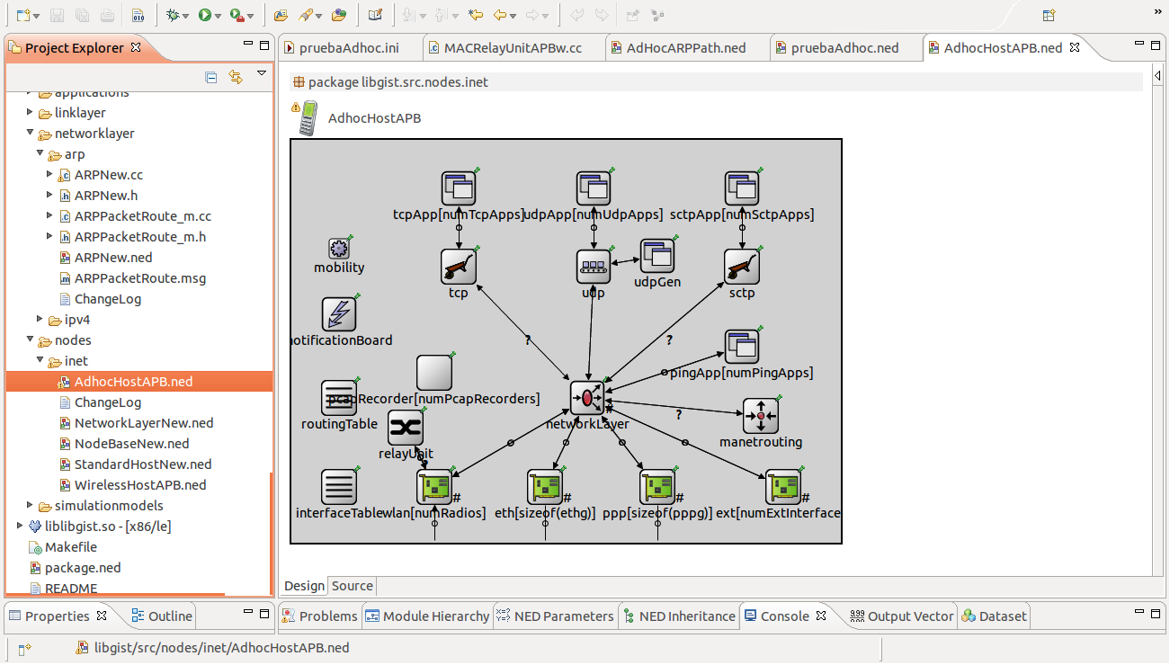}
\caption{Implementation of the \texttt{AdhocHostAPB} module in OMNeT++ 4.2.2 and INET 2.0.0}
\label{fig:adhochostapb}
\end{figure*}

As mentioned in the introduction, the ARP-Path protocol had been previously implemented in different platforms. However, most of them are intended for wired networks. Therefore, we decided to implement wARP-Path in the OMNeT++ simulator, which was the fastest alternative to check the suitability of the protocol for wireless networks.

\subsection{Implementation in OMNeT++ 4.2.2 and INET 2.0.0}
The wARP-Path was first developed by modifying some parts of the modules defined for ad hoc networks in the INET framework for OMNeT++, specifically the modification was done in the so-called \texttt{AdHocHost} module which was converted into the a new module called \texttt{AdHocHostAPB} by adding a relay submodule in it, as it can be seen in Fig.~\ref{fig:adhochostapb}.

The \texttt{AdHocHost} module implements, as it name recalls, a host for wireless ad hoc communications. This module is composed of several submodules. For instance, in the lower part, we can see several submodules directly related to the physical layer (\texttt{wlan}, \texttt{eth}, \texttt{ppp}, etc), while in the upper part there are modules associated to the application (\texttt{tcpApp}, \texttt{udpApp}, etc) and transport (\texttt{tcp}, \texttt{udp}, etc) layer. At the same time, most of the submodules are connected to the network layer submodule (called \texttt{networkLayer}), which is responsible of routing decisions and it can apply different protocols for it, such as AODV. 
The module developed for implementing the wARP-Path PoC was called \texttt{AdHocHostAPB}, which is shown in Fig.~\ref{fig:adhochostapb}, and it is an extension of the \texttt{AdHocHost}, which simply adds and intermediate module between the network layer and the physical layer, and it is called relayUnit. The relayUnit submodule applies learning and forwarding based on the frames received from the wlan submodule, specifically it applies the pseudocode shown before in Listing~\ref{pseudocode} for any frame received from wlan.

\subsection{Implementation in OMNeT++ 5.2 and INET 3.6.3}
%Hedayat: Could you please add a few lines just explaining the implications of implementing wARP-Path in the newest OMNeT++ version and what framework would be the most suitable for it? Thanks!

Lately, the wARP-Path protocol was reimplemented using the latest version of OMNeT++ and INET framework (code available in~\cite{WARP-Path}). The reason behind is that INET is quickly updated, and it covers a wide variety of protocols and components, and also, it is taken as a base for several other simulation frameworks~\cite{INET}.

Most of the differences in the two implementations are due to the evolution of the INET framework. In the recent versions of INET framework, the forwarding table (\texttt{macTable}) has been separated from the MAC relay unit (\texttt{MACRelayUnit}). Based on this, in this new implementation, the Learning/Lookup Table (\texttt{LT}) and Blocking/Broadcast Table (\texttt{BT}) are separated from the \texttt{MACRelayUnit} and placed beside the \texttt{MACRelayUnit} as two independent modules  to provide the service to the \texttt{MACRelayUnit}, as it can be seen in Fig.~\ref{fig:adhochostapb}. All steps of the algorithm excluding line 05 (as shown in Listing~\ref{pseudocode}) are implemented in the \texttt{Ieee80211MgmtAdhocAPB} module, and line 05 is implemented in the \texttt{MACRelayUnitWAPB} module.

\begin{figure*}[hbt]
\centering
\includegraphics[width=0.65\textwidth]{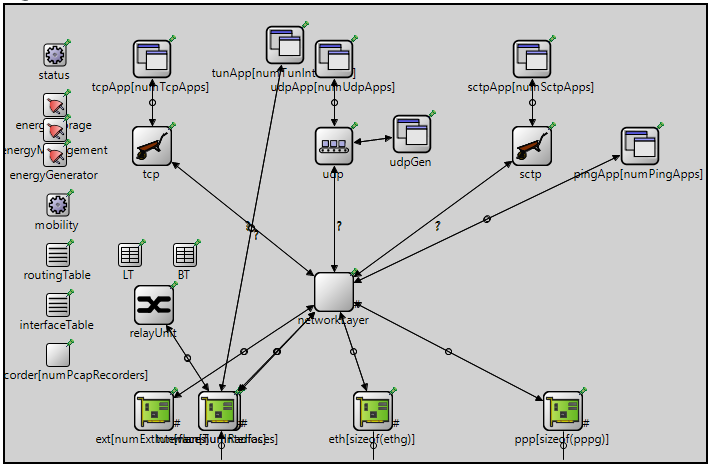}
\caption{Implementation of the \texttt{AdhocHostAPB} module in OMNeT++ 5.2 and INET 3.6.3}
\label{fig:adhochostapb}
\end{figure*}

\begin{figure*}
        \centering
        \begin{subfigure}[htb]{0.80\textwidth}
                \includegraphics[width=\textwidth]{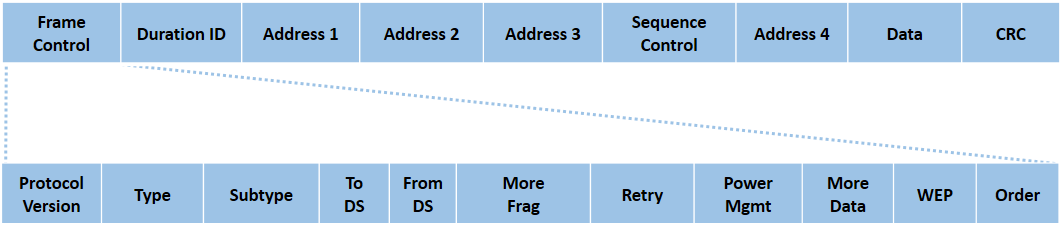}
                \caption{IEEE 802.11 frame format}
                \label{fig:adr:ieeeff}
        \end{subfigure}%
        \quad
        \begin{subfigure}[htb]{0.48\textwidth}
                \includegraphics[width=\textwidth]{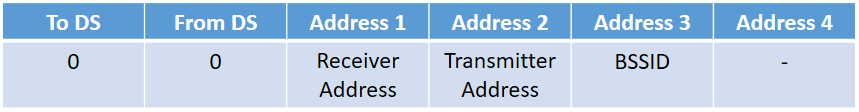}
                \caption{Interpretation of the MAC Addresses in the Ad-hoc mode
}
                \label{adhocadr}
        \end{subfigure}
        \begin{subfigure}[htb]{0.48\textwidth}
                \includegraphics[width=\textwidth]{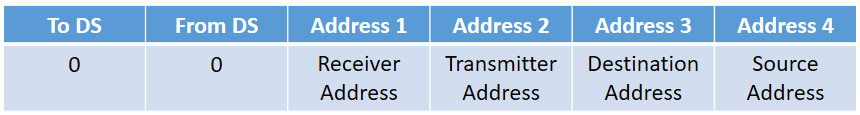}
                \caption{Interpretation of the MAC Addresses in wARP-Path
}
                \label{fig:adr:wap}
        \end{subfigure}
        \caption{Applying the address fields of IEEE 802.11 frame format in wARP-Path}\label{fig:adr}
\end{figure*}

There are two addresses for forwarding between two sequential hops (\textit{physical} addresses) and two other addresses to indicate the beginning and end of the path (\textit{logical} addresses). All four address fields embedded in the IEEE 802.11 MAC frame format (as shown in Fig.~\ref{fig:adr:ieeeff}) are used, which are receiver address as physical receiver, transmitter address as physical transmitter, destination address as logical receiver, and source address as logical transmitter respectively (as shown in Fig.~\ref{fig:adr:wap}). The physical addresses change in each hop, and the logical addresses do not change in the intermediate nodes, but in the final destination. To achieve the transparency required for the upper layer, the logical addresses are put in the physical address space before decapsulating the frame.

\subsection{Regarding OMNeT++/INET and their implementation of media access control in wireless networks}
In OMNeT++, frames are usually flooded at the same exact simulated time. Accordingly, when there is more than a single path to reach destination and the frame is being flooded for path exploration as in wARP-Path, it is high probability that some nodes will simultaneously broadcast this frame, e.g. an ARP Request. Since there is no unique receiver for sending the broadcast packets, no control packets (\texttt{Request-to-Send (RTS)} and \texttt{Clear-to-Send (CTS)}) are used and only carrier sense on the transmitter is performed because of the lack of coordination of the receivers and the possibility of a collision between the \texttt{CTSs}~\cite{MACAW}. If a broadcast packet collides, in addition to wasting a further capacity of the channel since the broadcast packets are larger than the control packets, it also has a negative impact on the quality of the explored paths. Therefore, a solution to this problem is necessary.

In the AODV implementation of INET framework, using a random jitter before sending protocol control packets such as \texttt{hello}, \texttt{Route Request (RREQ)}, and \texttt{Route Reply (RREP)} messages according to RFC 5148~\cite{jitter} has been adopted to overcome this problem. In this way, two random jitter has been used. First, when AODV generates the periodic \texttt{hello} messages  in INET framework, for this type of simultaneity in RFC 5148, a random value \texttt{(jitter)} is subtracted from the time interval between two consecutive transmission of the same type messages\texttt{(MESSAGE\_INTERVAL)}. Using subtract instead of sum prevents excessive delay in receiving messages. Second, when AODV generates the \texttt{RREQ} and \texttt{RREP} messages in INET framework, since these messages are not periodic messages, there is not any \texttt{(MESSAGE\_INTERVAL)} to calculate the delay. In this type of simultaneity, RFC 5148 introduces jitter in an interval between zero and \texttt{MAXJITTER}.

We use the same approach to overcome the simultaneously broadcast problem in wARP-Path, with the difference that the control messages in this protocol are the standard ARP packets. Therefore, the mechanisms mentioned above apply to these packets similarly.
%We introduce another approach based on IEEE802.11 to reducing the impact of this problem. When a broadcast packet occupies a channel, in order to reduce the possibility of sending another broadcast packets from its neighbors, we reduce the possibility of broadcasting the neighboring nodes as long as the neighbors have not received the interference of the packet. To reach this goal, we split the \texttt{Contention Window (CW)} into the duration of the broadcast packet, and back-off value is randomly obtained in this new range. 

%Although adoption of a big value for the back-off value can reduce the probability of collisions, on the other hand, it increases transmitter`s idle time, and channel capacity is wasted in terms of wasteful idle time~\cite{MACAW}. Also, adoption of a small value for the back-off value can increase the probability of collisions, and channel capacity is wasted in terms of wasteful collisions~\cite{MACAW}. Therefore, obtaining an optimal trade-off between the optimal use of channel capacity and throughput on the one hand, and the reduction of the probability of collision of packets on the other hand, is an issue we will consider. 

%Nevertheless, there are other problems such as the hidden and expose terminal for sending a broadcast packet since transmitter node does not have any information about the status of the interference and the collision at the receiver's nodes because of the lack of control packets. 

%% file: text/evaluation.tex
\section{Evaluation}
\label{evaluation}
This chapter is devoted to evaluate wARP-Path. An initial thought was to compare it with AODV, but we found a bug in the implementation of AODV in the INET framework \footnote{More specifically, some \texttt{RREQ} packets are deleted, and it has a negative impact on the end-to-end delay and quality of discovered routes of AODV.}. Therefore, we finally decided to exclude the evaluation of AODV (as results were not reliable) from our analysis.
%Also, since we experienced different results for running our simulation in \texttt{Graphical User Interface (GUI)} of OMNeT++ and INET framework (slow and express modes gave us different results), our results is according to \texttt{CMDENV} mode of this simulator. ==> based on some changes in flowgenerator such as using constant time (10 s) between each session, assigning a physical Random Number Generator to the flowGenerator module (by *.generator.rng-0 = 3 in *.ini file), this problem (different results) was resolved.

In this evaluation, we first define the test cases and scenarios. We then briefly compare wARP-Path and ARP-Path, and finally we analyze wARP-Path in terms of \emph{goodput ratio} and \emph{average end-to-end delay}.

\subsection{Definition of test cases and scenarios}
In order to evaluate the wARP-Path protocol, our model was inspired by~\cite{perkins1999aodv} and~\cite{broch1998performance}. We defined 50 nodes uniformly distributed within a fixed-size area of 1500 * 1500. To manage centralized behaviors such as: (1) the uniform election of a source and destination node between all nodes to start a session, (2) the selection of traffic for a session based on Table~\ref{table:traffic}, or (3) the computation of the average metrics for all nodes and flows in the network, we defined a module called \texttt{flowGenerator} in the simulation. Interval time of the simulation is 600s and the first session is started at time 0.2s. Next session is 
%uniformly started in the interval of [0s,18s) 
started after 10 seconds and other sessions are 
%uniformly started the same interval after starting the previous session.
started after the same time, periodically.
By using a fixed time (i.e. 10s) instead of a random time in an interval, we adopted enough interval between the flows so that we can analyze the effect of the ARP Request messages on other flows. The total number of sessions in the simulation is 10.

\begin{table}[!h]
\centering
\caption{Traffic Parameters}
\label{table:traffic}
\begin{tabular}{|l|l|l|}
\hline
\textbf{Parameter}& \textbf{S\_DATA}& \textbf{VOICE}\\
\hline
Transport protocol& UDP& UDP\\
Session interval& Geometric (mean 900)& Geometric (mean 600)\\
Packet size& 64 Bytes& 160 Bytes\\
Packet send interval & 20 ms& 20 ms\\
\hline
\end{tabular}
\end{table}

Two types of traffic will be analyzed: S\_DATA (that stands for \textit{small data}) and VOICE, as defined in~\cite{perkins1999aodv}. \\
To generate S\_DATA traffic, as it can be seen in Table~\ref{table:traffic}, data packets contain 64 bytes, and inter-arrival time of data packets is 20 ms. Therefore, this traffic is produced with a rate of 25,6 Kbps in each selected host as a source per session. \\
To generate VOICE traffic, as also illustrated in Table~\ref{table:traffic}, we suppose that quality of voice is telephony, so sampling frequency is 8 KHz. Since each sample is expressed with 1 byte, each selected host as a source per session produces a traffic with rate of 64 Kbps. Given that each packet is generated every 20 ms, the packet size is 160 bytes.

Each node has one transmitter and one receiver, and all nodes use the same channel model. Their physical and MAC layers are based on IEEE 802.11~\cite{ieee80211}, considering the \emph{Distributed Coordination Function (DCF)} mode, whose properties are shown in Table~\ref{table:simulation}. 

\begin{table}[!h]
\centering
\caption{Simulation Parameters}
\label{table:simulation}
\begin{tabular}{|l|l|}
\hline
\textbf{Parameter}& \textbf{Value}\\
\hline
Simulation time& 600 s\\
Traffic generation start time& 0.2 s \\
Traffic generation start time& 600 s\\
\hline
\multicolumn{2}{|c|}{\textbf{wARP-Path}}
\\\hline 
LT aging time& 120 s\\
BT blocking time& 1 s\\
MAXJITTER& 5 ms\\
Jitter& uniform(0, MAXJITTER)\\
\hline
\multicolumn{2}{|c|}{\textbf{ARP protocol}}
\\\hline 
ARP retry count& 5\\
ARP retry timeout& 200 ms\\
ARP cache timeout& 120 s\\
\hline
\multicolumn{2}{|c|}{\textbf{MAC layer}}
\\\hline 
IEEE802.11 type& IEEE802.11 g\\
Maximum size of queue& 14\\
MAC retry limit& 7\\
CW min (for S\_DATA)& 15 time slots\\
CW min (for VOICE)& 20 time slots\\
\hline
\multicolumn{2}{|c|}{\textbf{Phy layer}}
\\\hline 
Carrier frecuency& 2.4 GHz\\
Bandwidth& 2 MHz\\
Modulation scheme& BPSK + DSSS\\
Bit rate& 1 Mbps\\
Transmit Power& 2 mW\\
Receiver sensitivity& -85 dBm\\
SINR threshold& 4dB\\
Energy detection threshold& -85 dBm\\
Path loss type& Free Space Path Loss\\
Background noise power& -110 dBm\\
\hline

%Network size& 1500m * 1500m\\ % it is commented in context
%Number of nodes& 50\\ % it is commented in context
%Distribution of nodes& Uniform\\ % it is commented in context
\end{tabular}
\end{table}

\subsection{Comparison with ARP-Path}
The wARP-Path protocol is quite similar to ARP-Path. But in wARP-Path, since intermediate nodes are hosts (not only switches) that can create a new session to each destination, they can use the paths explored by each node that is not an intermediate node. According to computation of forwarding state in the ARP-Path protocol~\cite{Rojas15}, in the worst case of ARP-Path, when all nodes (non-intermediate) communicated with each other, the paths between some non-intermediate nodes and some intermediate nodes had been creating. Now, in wARP-Path, since intermediate nodes have the capability to start a session with non-intermediate nodes, they can start some additional sessions without adding new entries in the forwarding tables, and this amount of communications without inserting a new entry in tables is a payoff of wARP-Path against ARP-Path protocol. Briefly, with the same number of entries in forwarding tables, wARP-Path might have a higher number of active communications than ARP-Path.

\subsection{Goodput Ratio}
%\subsubsection{Goodput ratio}
%\textbf{Goodput ratio} \\

\begin{figure*}
        \centering
        \begin{subfigure}[htb]{0.80\textwidth}
                \includegraphics[width=\textwidth] {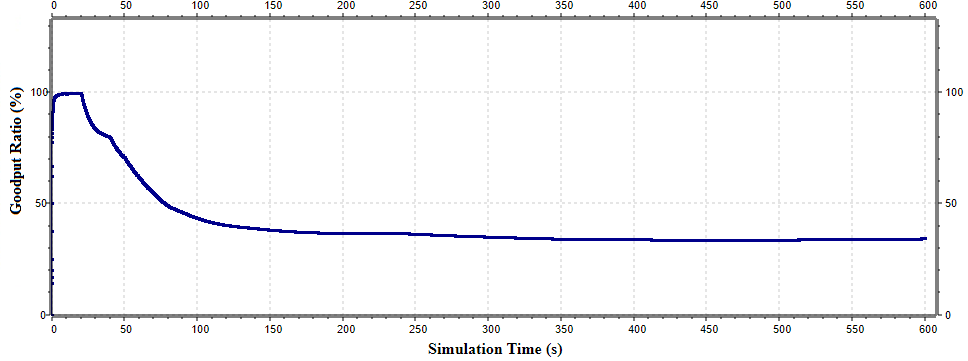}
                \caption{S\_DATA}
                \label{fig:good:sdata}
       \end{subfigure}%
        \quad
        \begin{subfigure}[htb]{0.80\textwidth}
                \includegraphics[width=\textwidth]{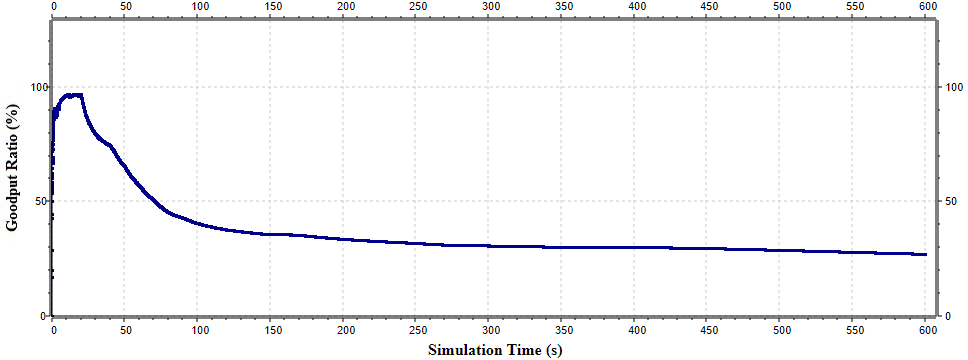}
                \caption{VOICE}
                \label{fig:good:voice}
        \end{subfigure}       
        \caption{Achieved Goodput Ratio}\label{fig:good}
\end{figure*}

\begin{table*}[!h]
\centering
\caption{Generated flows}
\label{table:flows}
\begin{tabular}{|l|l|l|l|l|l|l|l|c|}
\hline
\multicolumn{9}{|c|}{\textbf{S\_DATA Traffic}}
\\\hline 
\textbf{Session \#}& \textbf{Source}& \textbf{Destination}& \textbf{Amount of Traffic}& \textbf{Start Time}& \textbf{ End Time}& \textbf{\# ARP Requests}& \textbf{\# ARP Replies}& \textbf{Path Discovered}\\
\hline
1& Host 42& Host 24& 3.5424e+006 B& 0.2 s& 1107.2 s& 163& 19& \checkmark\\
2& Host 3& Host 8& 2.0576e+006 B& 10.2 s& 653.2 s& 5& 5& \checkmark\\
3& Host 21& Host 19& 6.5216e+006 B& 20.2 s& 2058.2 s& 3165 (flows 3, 5)& 0& -\\
4& Host 43& Host 41& 665600 B& 30.2 s& 238.2 s& 2& 2& \checkmark\\
5& Host 21& Host 38& 1.6704e+006 B& 40.2 s& 562.2 s& 3165 (flows 3, 5)& 29& \checkmark\\
6& Host 20& Host 44& 451200 B& 50.2 s& 191.2 s& 51& 7& \checkmark\\
7& Host 39& Host 14& 3.0144e+006 B& 60.2 s& 1002.2 s& 71& 16& \checkmark\\
8& Host 26& Host 17& 1.2544e+006 B& 70.2 s& 462.2 s& 2000& 0& -\\
9& Host 22& Host 2& 2.5696e+006 B& 80.2 s& 883.2 s& 6& 6& \checkmark\\
10& Host 1& Host 26& 860800 B& 90.2 s& 359.2 s& 1375& 0& -\\
\hline
\multicolumn{9}{|c|}{\textbf{VOICE Traffic}}
\\\hline
\textbf{Session \#}& \textbf{Source}& \textbf{Destination}& \textbf{Amount of Traffic}& \textbf{Start Time}& \textbf{ End Time}& \textbf{\# ARP Requests}& \textbf{\# ARP Replies}& \textbf{Path Discovered}\\
\hline
1& Host 42& Host 24& 5.904e+006 B& 0.2 s& 738.2 s& 18& 5& \checkmark\\
2& Host 3& Host 8& 3.424e+006 B& 10.2 s& 438.2 s& 2& 2& \checkmark\\
3& Host 21& Host 19& 1.0864e+007 B& 20.2 s& 1378.2 s& 2818 (flows 3, 5)& 0& -\\
4& Host 43& Host 41& 1.112e+006 B& 30.2 s& 169.2 s& 1& 1& \checkmark\\
5& Host 21& Host 38& 2.784e+006 B& 40.2 s& 388.2 s& 4 (flows 3, 5)& 4& \checkmark\\
6& Host 20& Host 44& 752000 B& 50.2 s& 144.2 s& 16& 3& \checkmark\\
7& Host 39& Host 14& 5.016e+006 B& 60.2 s& 687.2 s& 54& 8& \checkmark\\
8& Host 26& Host 17& 2.088e+006 B& 70.2 s& 331.2 s& 525& 0& -\\
9& Host 22& Host 2& 4.28e+006 B& 80.2 s& 615.2 s& 2& 2& \checkmark\\
10& Host 1& Host 26& 1.432e+006 B& 90.2 s& 269.2 s& 360& 0& -\\
\hline
\end{tabular}
\end{table*}

The goodput ratio is calculated over the entire network when a host generates or receives a packet on the network. Therefore, goodput ratio is a function of time in the simulation time interval, and we denote it as follows:

$${GoodputRatio(t)} = {\frac {n^r_t} {n^g_t} \times 100}$$ \\
where $n^g_t$ is the number of bytes in packets generated by the application layer of hosts from the start time of the simulation until time $t$, and $n^r_t$ is the number of bytes in packets received by the application layer of hosts from the start time of simulation until time $t$. The reason of using number of bytes instead of the number of packets is because the last packet of each session might vary in size, depending of the session, although this situation did not occur in our traffic.

Figures~\ref{fig:good:sdata} and~\ref{fig:good:voice} respectively show the achieved goodput ratio as a function of time for S\_DATA and VOICE traffic with 10 sessions. As depicted in the figures, at the initial moments of the simulation, goodput ratio is low since source hosts only generated packets, and no destination host received packets. After this small interval, once destinations received packets, the goodput ratio increasingly grows near to 100\%. This situation is stable until the number of sessions and broadcasts increase. \\
Although using \emph{jitter} is a good approach to overcome broadcast problems, this problem can still cause collision with other broadcast (simultaneously forwarding), \emph{RTS} (simultaneously forwarding), \emph{CTS} (when the \emph{RTS} sender is hidden from broadcast sender), or even \emph{ACK} or \emph{data} (when \texttt{CTS} has collided and a node that is hidden from the \texttt{RTS} sender sends a broadcast) packets in dense and high broadcast scenarios. Therefore, when the number of broadcasts increases in the network, as mentioned in previous section, it negatively affects on the channel capacity and quality of the explored paths (and even a path can not be discovered though it exists), which causes reduction of the goodput ratio. \\
Time interval between two consecutive packets is the same in both traffic types, but since VOICE packets are bigger than S\_DATA packets, the probability of collision increases with VOICE packets, and the loss of a packet has a greater impact on its respective goodput ratio.

The traffic is \texttt{UDP}, connection-less and unreliable transport protocol, and there is no hand shaking, so traffic is continuously injected to lower layer. As shown in Table~\ref{table:flows} (i.e. sessions 3, 8, and 10 in both traffic types), a lot of traffic is wasted before a path is found (or there is no path at all), which causes that the goodput ratio stays at the same low value. For S\_DATA traffic, they produce 3970560 bytes until the end of simulation. If we reduce this value (i.e. total sent bytes in Table~\ref{table:results}), the goodput ratio is increased up to 48.12\%. 

As shown in Table~\ref{table:flows}, another impact of the injected traffic to lower layers are the ARP Request messages, which are frequently sent to the network. Collisions resulting from these broadcasts will negatively affect the goodput ratio and the delay. Collision of the broadcast packets with control and data packets will cause the MAC layer to increase CW and select a back-off value in a larger range. Therefore, delay increases in the interval in which ARP Request messages enter the network, and also using the jitter will cause a delay in starting a transmission. %\\

%Other factor which negatively affects Goodput Ratio is the delay in discovering a path. since one packet is produced each 20 ms, in order to avoid dropping packets due to the queue filling before discovering a path, Based on the parameters of ARP Protocol shown in~\ref{table:simulation}, an appropriate size queue is required:
%$$Len_Q = 1 + {ARP Retry Count} \times \frac{ARP Retry Timeout}{data Send Interval}$$
%where ${ARP Retry Count} \times \frac{ARP Retry Timeout}{Send Interval of data packets}$ is the number of packets waited for the ARP module before discovering a path in the worst case. ==> input rate and service rate ...

%Other factor => each 120s arp cache is time outed, new arp request results loss  

%Because of this, ARP requests are frequently entered in to the network and cause that Goodput Ratio stays at the same low value. ==> if ARP is entered frequently, why is the delay decreased after that peak? (response: because 3970560 bytes, this is in addition to ARP effects) next action ==> evaluation of ARP, specification (source and destination, ..) of discovered paths(same as the path optimality metric),

%AODV~\cite{perkins1999aodv} and ~\cite{broch1998performance}

\subsection{End-to-end Delay}
%\subsubsection{Delay}
%\textbf{Delay} \\

\begin{figure*}
        \centering
        \begin{subfigure}[htb]{0.80\textwidth}
                \includegraphics[width=\textwidth]{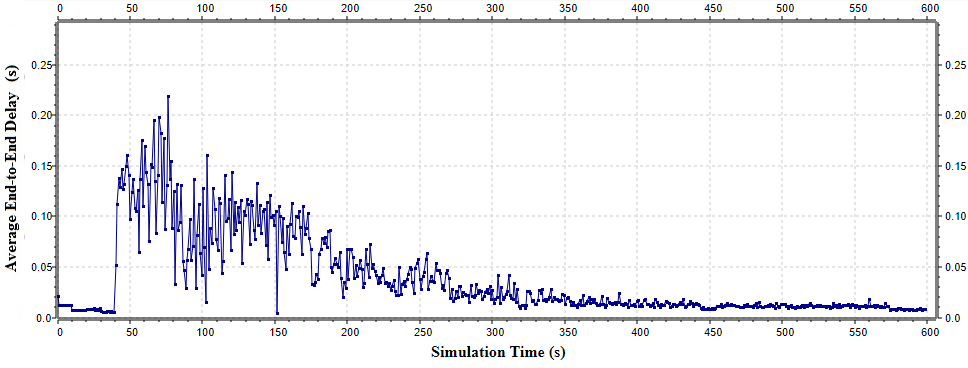}
                \caption{Average end-to-end delay computed in each interval for all hosts}
                \label{fig:e2e:S-DATA:invl}
        \end{subfigure}%
        \quad
        \begin{subfigure}[htb]{0.80\textwidth}
                \includegraphics[width=\textwidth]{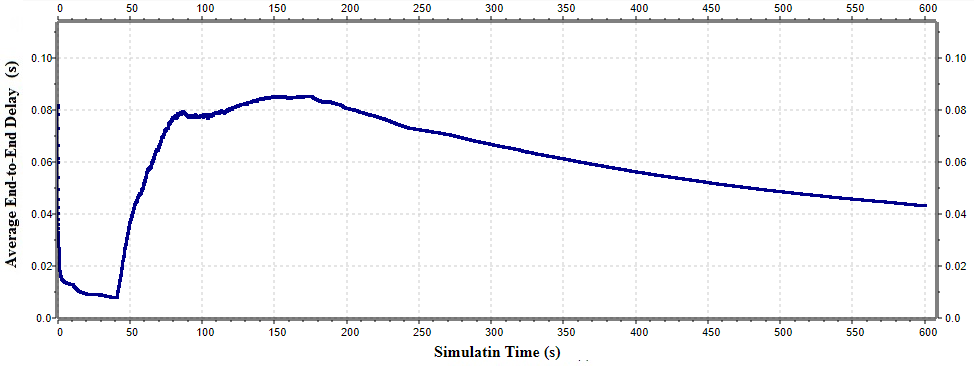}
                \caption{Average end-to-end delay of all hosts}
                \label{fig:e2e:S-DATA:all}
        \end{subfigure}
        \quad
        \begin{subfigure}[htb]{0.80\textwidth}
                \includegraphics[width=\textwidth]{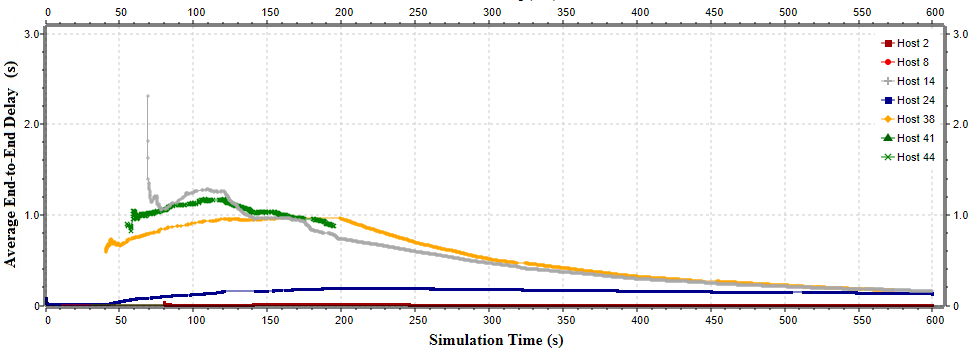}
                \caption{Average end-to-end delay of each host}
                \label{fig:e2e:S-DATA:host}
        \end{subfigure}
        
        \caption{Achieved end-to-end delay for S\_DATA traffic}\label{fig:e2e:S-DATA}
\end{figure*}

\begin{figure*}
        \centering
        \begin{subfigure}[htb]{0.80\textwidth}
                \includegraphics[width=\textwidth]{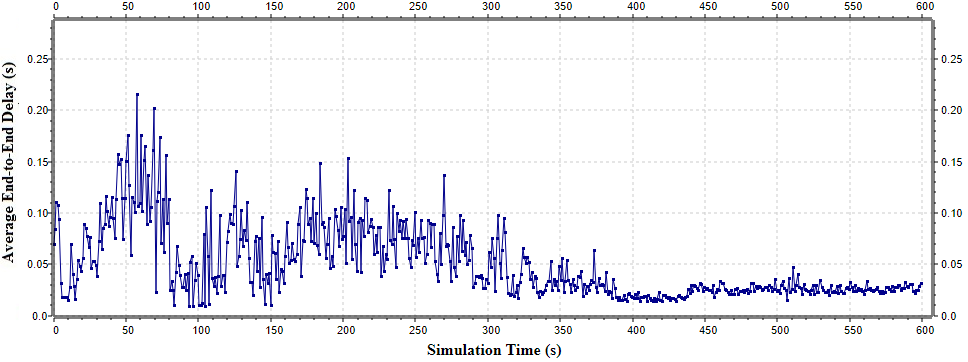}
                \caption{Average end-to-end delay computed in each interval for all hosts}
                \label{fig:e2e:VOICE:invl}
        \end{subfigure}%
        \quad
        \begin{subfigure}[htb]{0.80\textwidth}
                \includegraphics[width=\textwidth]{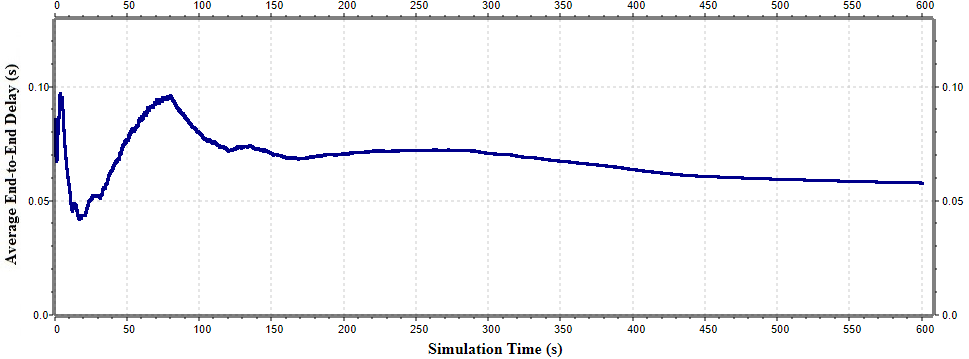}
                \caption{Average end-to-end delay of all hosts}
                \label{fig:e2e:VOICE:all}
        \end{subfigure}
        \quad
        \begin{subfigure}[htb]{0.80\textwidth}
                \includegraphics[width=\textwidth]{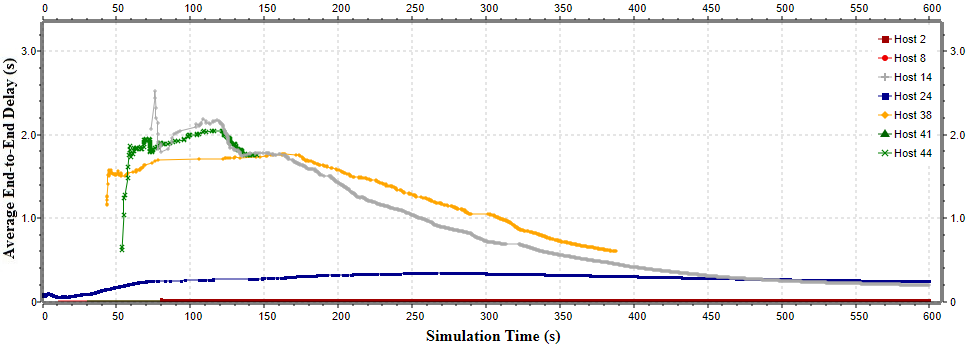}
                \caption{Average end-to-end delay of each host}
                \label{fig:e2e:VOICE:host}
        \end{subfigure}
        
        \caption{Achieved end-to-end delay for VOICE traffic}\label{fig:e2e:VOICE}
\end{figure*}

\begin{table}[!h]
\centering
\caption{Results at the end of simulation}
\label{table:results}
\begin{tabular}{|l|l|l|}
\hline
& \textbf{S\_DATA}& \textbf{VOICE}\\
\hline
\textbf{Goodput ratio}& 34.22\%& 26.99\%\\
\textbf{Average end-to-end delay}& 0.0431 s& 0.0579 s\\
\textbf{Average end-to-end delay last intvl}& 0.0081 s& 0.0316 s\\
\textbf{Total sent bytes}& 1.3742848E7& 2.95056E7\\
\textbf{Total sent packets}& 214732& 184410\\
\textbf{Total received bytes}& 4702784& 7964160\\
\textbf{Total received packets}& 73481& 49776\\
\hline
\end{tabular}
\end{table}

Additionally, we computed the average end-to-end delay of the network based on the following equation for both each host and all possible destinations in the network.

$${\overline {Delay_{e2e}}(t)} = {\frac {\sum_{i\in \{p \mid t_{r_p} \le t\}} d_i} {n_{set}}}, {d_i = t_{r_i} - t_{g_i}}$$

Considering the equation, $p$ is a packet received to the host for which we want to calculate average end-to-end delay, $t_{r_p}$ is the arrival time of packet $p$ at the application layer, $d_i$ denotes the end-to-end delay of packet $i$ which is obtained by subtracting the arrival time of the packet ($t_{r_i}$) from the generation time of the packet ($t_{g_i}$), and ${n_{set}}$ is the number of elements in the set. In case of calculating the average end-to-end delay of all the network, $p$ is the packet received by one host.

Figures~\ref{fig:e2e:S-DATA:host} and~\ref{fig:e2e:VOICE:host} respectively show the average end-to-end delay of each host for the traffics of S\_DATA and VOICE for 10 sessions. Figures~\ref{fig:e2e:S-DATA:all} and~\ref{fig:e2e:VOICE:all} respectively show the average end-to-end delay of all the network for the traffics of S\_DATA and VOICE with 10 sessions.

The values just mentioned give us good information about the overall and current state of the network. To calculate end-to-end delay, since there are peaks at the end-to-end delay of the network and in order to balance these peaks, we obtain the average end-to-end delay in the small intervals (for example, this interval we will denoted by $\Delta t$, and it is 1 second in our simulation) based on the following equation, instead of pure end-to-end delay (as shown in Figures~\ref{fig:e2e:S-DATA:invl} and~\ref{fig:e2e:VOICE:invl}). 

$${\overline {IntervalDelay_{e2e}}(t)} = {\frac {\sum_{i\in \{p \mid \Delta t \lfloor \frac {t}{\Delta t} \rfloor \le t_{r_p} < \Delta t \lfloor \frac {t}{\Delta t} \rfloor + \Delta t\}} d_i} {n_{set}}}$$
$$, {d_i = t_{r_i} - t_{g_i}}$$

%% file: text/discussion.tex
\section{Discussion}
\label{discussion}

Finally, we discuss different aspects of the wARP-Path protocol. Particularly, in some of the topics we compare wARP-Path to AODV, which we consider the most similar protocol, as they are both reactive routing protocols for layer 2 and layer 3, respectively. \\

\textbf{Layer 2 vs. Layer 3} \\
wARP-Path acts on Layer 2, and since this protocol tries to find the shortest paths with the least load and delay, so it can as well take advantages of this layer on the network, such as higher speed and lack of processing latency due to layer 3 routing. \\
Since on-demand routings are based on query reply, they endure a delay to find a route~\cite{Iwata99}. However, ARP-Path uses standard ARP Request and ARP Reply packets for exploring a path, it does not include this delay. \\

\textbf{Scalability} \\
%Regarding scalability, on one hand, AODV requires specific messages for routing and, moreover, it needs a modification of the network layer to be implemented. On the other, wARP-Path seems to have a worse scalability than AODV in terms of table size, as AODV can merge entries (based on IP), while wARP-Path cannot.
In on-demand protocols, when a node wants to communicate with a destination, the route is computed. Therefore, they do not store route information for all destinations permanently, so this feature increases their scalability to be used in large networks~\cite{Iwata99}. Despite this common feature in both wARP-Path and AODV, each protocol has unique features which cause differences in their scalability. In spite of the mentioned advantages of Layer 2 for wARP-Path, this protocol uses the flat addressing structure in layer 2, and AODV uses the hierarchical structure of IP addressing in Layer 3. Hierarchical routing increasingly reduces the size of routing tables and processing overhead~\cite{Iwata99}, whereas using the flat addressing structure raises the problem of increasing the number of entries in the forwarding tables. \\
In general, we found that scalability was the first issue when migrating the ARP-Path protocol to wireless networks. \\
%Scalability (larger coverage area implies repetition.......)
%Tables... (sensing devices)

\textbf{Stability vs Mobility} \\
Wired networks are more stable than wireless networks due to their nature in using fixed nodes and links. This is one of the reasons why the protocols used in these two types of networks are different. Since the All-Path protocols have been designed for wired networks, they are inherently stable. ARP-Path, as one of these protocols, tries to maintain the stability of the path until the end of each communication unless network physical stability is lost. In this case, by sending the path recovery messages, it starts to discover the path in the unstable parts of the network. The stability of paths is maintained by refreshing the lifetime of the entries in the forwarding tables until the end of a communication. \\
But in AODV, designed for using wireless networks, this bound of stability seen in the ARP-Path is not seen here. The discovered routes in AODV are not necessarily kept to the end of communication, and life time of the routes in the routing tables are not updated by transferring data packets, they are updated only by transferring protocol packets such as RREQ and RREP. The existence of this feature in AODV increases its efficiency to overcome the unstable state of nodes and links in wireless networks. \\

\textbf{Path repair} \\
Another issue found when migrating the protocol was path repair. In wired networks, link failure detection is direct, as usually the physical layer provides this feature. However, link failure detection in wireless networks requires additional mechanisms, not only to probe if neighbors are still available, but also to guarantee if packets are effectively reaching their destinations. \\
Additionally, the path repair mechanism in ARP-Path requires broadcasting and might be too costly for wireless networks, especially when their nodes have mobility. So simple methods such as broadcasting might be more efficient. That is the main reason why we did not implement and test path repair in wARP-Path. \\

\textbf{Wireless frame format} \\
ARP-Path leverages the fact that the most common frame format in wired networks is Ethernet. However, in wireless, there is a wide range of layer 2 frames and protocols, such as WiFi, WiFi-Direct, Bluetooth, Bluetooth SMART or BLE~\cite{Gomez12}, LR-WPAN (802.15.4) or Zigbee. \\
This diversity affects the implementation of wARP-Path, which might have variations depending on the layer 2 implemented. \\

\textbf{Software-Defined Networking} \\
Software-Defined Networking (SDN)~\cite{Kreutz15} is flourishing rapidly and, although the initial deployments were based on wired networks, wireless networks are also targeted as part of the SDN spectrum, which makes harder the appearance of new distributed protocols. \\
Thus, wARP-Path is not good enough to beat the advantages of SDN and a more disruptive approach to wireless networks should be applied, instead of a simple migration of a protocol from wired to wireless.

%\subsubsection{Routing overhead}
%~\cite{broch1998performance}

%\subsubsection{Path optimality}
%~\cite{broch1998performance}

%% file: text/conclusion.tex
\section{Conclusion}
\label{conclusion}

Along the article we have studied the implications of adapting a wired bridging protocol to a wireless environment. Apart from the specificities of the implementation, we have discovered to main drawbacks during the migration: 
\begin{enumerate}
	\item \textbf{Frame format}: Wired bridging protocols are mainly Ethernet-based, while in wireless the frame format is diverse and we had to choose one to continue the implementation (i.e. moving from one frame format to other might not be necessarily straightforward). This affects our protocol as is based on layer 2, but it would not affect layer 3 routing protocols.
    \item \textbf{Flooding and scalability}: While broadcast in wired networks might be relatively useful in some scenarios, in wireless networks might be totally unacceptable. More specifically, wireless networks always flood the information per se (the radio signal is received by all nodes in the range), so adding an overhead to the forwarding protocol should imply --at least-- saving time in processing the frame, i.e. avoiding broadcasting the frame whenever possible. 
\end{enumerate}

Additionally, some mechanisms such as path repair are not applicable to wireless networks, which imply redesigning parts of the protocol (e.g. to define some type of keep-alive or mobility-awareness mechanism).

Therefore, although ARP-Path is a simple and efficient protocol for wired bridging, wARP-Path did not show the equivalent benefits for wireless networks. The main conclusion is that wireless bridging protocols should be more efficient than routing protocols in different aspects (for example, drastically decreasing table size), with a groundbreaking approach, otherwise protocols as AODV or even applying SDN might still be more suitable for wireless networks.

%% file: wAP.bbl
\begin{thebibliography}{10}
\providecommand{\url}[1]{#1}
\csname url@samestyle\endcsname
\providecommand{\newblock}{\relax}
\providecommand{\bibinfo}[2]{#2}
\providecommand{\BIBentrySTDinterwordspacing}{\spaceskip=0pt\relax}
\providecommand{\BIBentryALTinterwordstretchfactor}{4}
\providecommand{\BIBentryALTinterwordspacing}{\spaceskip=\fontdimen2\font plus
\BIBentryALTinterwordstretchfactor\fontdimen3\font minus
  \fontdimen4\font\relax}
\providecommand{\BIBforeignlanguage}[2]{{%
\expandafter\ifx\csname l@#1\endcsname\relax
\typeout{** WARNING: IEEEtran.bst: No hyphenation pattern has been}%
\typeout{** loaded for the language `#1'. Using the pattern for}%
\typeout{** the default language instead.}%
\else
\language=\csname l@#1\endcsname
\fi
#2}}
\providecommand{\BIBdecl}{\relax}
\BIBdecl

\bibitem{Rojas15}
\BIBentryALTinterwordspacing
E.~Rojas, G.~Ib{\'a}{\~n}ez, J.~M. Gimenez-Guzman, J.~A. Carral,
  A.~Garcia-Martinez, I.~Martinez-Yelmo, and J.~M. Arco, ``{All-Path bridging:
  Path exploration protocols for data center and campus networks},''
  \emph{Computer Networks}, vol.~79, pp. 120 -- 132, 2015. [Online]. Available:
  \url{http://www.sciencedirect.com/science/article/pii/S1389128615000055}
\BIBentrySTDinterwordspacing

\bibitem{tcppath}
J.~Alvarez-Horcajo, D.~Lopez-Pajares, J.~M. Arco, J.~A. Carral, and
  I.~Martinez-Yelmo, ``{TCP-path: Improving load balance by network
  exploration},'' in \emph{2017 IEEE 6th International Conference on Cloud
  Networking (CloudNet)}, Sept 2017, pp. 1--6.

\bibitem{IbanezCL}
G.~Ib{\'a}{\~n}ez, J.~A. Carral, J.~M. Arco, D.~Rivera, and A.~Montalvo,
  ``{ARP-Path: ARP-Based, Shortest Path Bridges},'' \emph{Communications
  Letters, IEEE}, vol.~15, no.~7, pp. 770--772, 2011.

\bibitem{Ibanez10}
G.~Iba\~{n}ez, J.~Naous, E.~Rojas, D.~Rivera, J.~A. Carral, and J.~M. Arco,
  ``{A Simple, Zero Configuration, Low Latency Bridging Protocol},'' \emph{IEEE
  LCN Demos}, 2010.

\bibitem{Allan12}
D.~Allan and N.~Bragg, \emph{{802.1 aq Shortest Path Bridging Design and
  Evolution: The Architect's Perspective}}.\hskip 1em plus 0.5em minus
  0.4em\relax John Wiley \& Sons, 2012.

\bibitem{RBridges}
\BIBentryALTinterwordspacing
R.~Perlman, D.~E. Eastlake, D.~G. Dutt, S.~Gai, and A.~Ghanwani, ``{Routing
  Bridges (RBridges): Base Protocol Specification},'' RFC 6325, Jul. 2011.
  [Online]. Available: \url{https://rfc-editor.org/rfc/rfc6325.txt}
\BIBentrySTDinterwordspacing

\bibitem{Perlman11}
R.~Perlman, ``{Introduction to TRILL},'' \emph{The Internet Protocol Journal},
  vol.~4, no.~3, pp. 2--20, 2011.

\bibitem{aodv}
\BIBentryALTinterwordspacing
S.~R. Das, C.~E. Perkins, and E.~M. Belding-Royer, ``{Ad hoc On-Demand Distance
  Vector (AODV) Routing},'' RFC 3561, Jul. 2003. [Online]. Available:
  \url{https://rfc-editor.org/rfc/rfc3561.txt}
\BIBentrySTDinterwordspacing

\bibitem{aodv-paper}
\BIBentryALTinterwordspacing
E.~M. Belding-Royer and C.~E. Perkins, ``{Evolution and future directions of
  the ad hoc on-demand distance-vector routing protocol},'' \emph{Ad Hoc
  Networks}, vol.~1, no.~1, pp. 125 -- 150, 2003. [Online]. Available:
  \url{http://www.sciencedirect.com/science/article/pii/S1570870503000167}
\BIBentrySTDinterwordspacing

\bibitem{Iwata99}
A.~Iwata, C.-C. Chiang, G.~Pei, M.~Gerla, and T.-W. Chen, ``{Scalable routing
  strategies for ad hoc wireless networks},'' \emph{IEEE Journal on Selected
  Areas in Communications}, vol.~17, no.~8, pp. 1369--1379, Aug 1999.

\bibitem{dsr}
\BIBentryALTinterwordspacing
Y.-C. Hu, D.~A. Maltz, and D.~B. Johnson, ``{The Dynamic Source Routing
  Protocol (DSR) for Mobile Ad Hoc Networks for IPv4},'' RFC 4728, Feb. 2007.
  [Online]. Available: \url{https://rfc-editor.org/rfc/rfc4728.txt}
\BIBentrySTDinterwordspacing

\bibitem{Das00}
S.~R. Das, C.~E. Perkins, and E.~M. Royer, ``Performance comparison of two
  on-demand routing protocols for ad hoc networks,'' in \emph{Proceedings IEEE
  INFOCOM 2000. Conference on Computer Communications. Nineteenth Annual Joint
  Conference of the IEEE Computer and Communications Societies (Cat.
  No.00CH37064)}, vol.~1, 2000, pp. 3--12 vol.1.

\bibitem{He10}
L.~He, J.~Huang, and F.~Yang, ``{A noval hybrid wireless routing protocol for
  WMNs},'' in \emph{2010 International Conference on Electronics and
  Information Engineering}, vol.~1, Aug 2010, pp. V1--281--V1--285.

\bibitem{Zhu13}
L.~Zhu, C.~Lin, K.~Meng, and Y.~Dong, ``{P-AODV: A protection routing mechanism
  in wireless mesh networks},'' in \emph{2013 15th IEEE International
  Conference on Communication Technology}, Nov 2013, pp. 505--510.

\bibitem{Hong15}
\BIBentryALTinterwordspacing
L.~Hong, X.~Liu, L.~Zhang, and W.~Chen, ``{Towards sensitive link quality
  prediction in ad hoc routing protocol based on grey theory},'' \emph{Wireless
  Networks}, vol.~21, no.~7, pp. 2315--2325, Oct 2015. [Online]. Available:
  \url{https://doi.org/10.1007/s11276-015-0918-z}
\BIBentrySTDinterwordspacing

\bibitem{Suliman04}
I.~M. Suliman, T.~Saarinen, T.~M. Hautala, W.~Cheng, and T.~Braysy, ``{A
  comparison study between wireless bridging and routing},'' in
  \emph{International Workshop on Wireless Ad-Hoc Networks, 2004.}, May 2004,
  pp. 140--144.

\bibitem{Maurina10}
S.~Maurina, J.~Fitzpatrick, L.~Trifan, and L.~Murphy, ``{An enhanced
  bridged-based multi-hop wireless network implementation},'' in \emph{2010 The
  5th Annual ICST Wireless Internet Conference (WICON)}, March 2010, pp. 1--9.

\bibitem{ieee80211}
\BIBentryALTinterwordspacing
``{ IEEE 802.11TM WIRELESS LOCAL AREA NETWORKS}.'' [Online]. Available:
  \url{http://www.ieee802.org/11/}
\BIBentrySTDinterwordspacing

\bibitem{WARP-Path}
\BIBentryALTinterwordspacing
``{wARP-Path}.'' [Online]. Available:
  \url{https://github.com/gistnetserv-uah/AllPath-OMNeT/}
\BIBentrySTDinterwordspacing

\bibitem{INET}
\BIBentryALTinterwordspacing
``{INET Framework}.'' [Online]. Available: \url{https://inet.omnetpp.org/}
\BIBentrySTDinterwordspacing

\bibitem{MACAW}
V.~Bharghavan, A.~Demers, S.~Shenker, and L.~Zhang, ``{MACAW: a media access
  protocol for wireless LAN's},'' \emph{ACM SIGCOMM Computer Communication
  Review}, vol.~24, no.~4, pp. 212--225, 1994.

\bibitem{jitter}
\BIBentryALTinterwordspacing
T.~Clausen, C.~Dearlove, and B.~Adamson, ``{Jitter Considerations in Mobile Ad
  Hoc Networks (MANETs)},'' RFC 5148, 2008. [Online]. Available:
  \url{https://rfc-editor.org/rfc/rfc5148.txt}
\BIBentrySTDinterwordspacing

\bibitem{perkins1999aodv}
C.~Perkins and E.~Royer, ``Ad-hoc on-demand distance vector routing,'' in
  \emph{Proceedings Of the 2 IEEE workshop on Mobile Computing System and
  Applications}.\hskip 1em plus 0.5em minus 0.4em\relax IEEE, 1999, pp.
  90--100.

\bibitem{broch1998performance}
J.~Broch, D.~A. Maltz, D.~B. Johnson, Y.-C. Hu, and J.~Jetcheva, ``A
  performance comparison of multi-hop wireless ad hoc network routing
  protocols,'' in \emph{Proceedings of the 4th annual ACM/IEEE international
  conference on Mobile computing and networking}.\hskip 1em plus 0.5em minus
  0.4em\relax ACM, 1998, pp. 85--97.

\bibitem{Gomez12}
\BIBentryALTinterwordspacing
C.~Gomez, J.~Oller, and J.~Paradells, ``Overview and evaluation of bluetooth
  low energy: An emerging low-power wireless technology,'' \emph{Sensors},
  vol.~12, no.~9, pp. 11\,734--11\,753, 2012. [Online]. Available:
  \url{http://www.mdpi.com/1424-8220/12/9/11734}
\BIBentrySTDinterwordspacing

\bibitem{Kreutz15}
D.~Kreutz, F.~M.~V. Ramos, P.~E. Verissimo, C.~E. Rothenberg, S.~Azodolmolky,
  and S.~Uhlig, ``{Software-Defined Networking: A Comprehensive Survey},''
  \emph{Proceedings of the IEEE}, vol. 103, no.~1, pp. 14--76, Jan 2015.

\end{thebibliography}
